\documentclass[12pt]{article}
\usepackage{amsmath}
\usepackage{amssymb}
\usepackage{epsfig}

\begin{document}

\begin{center}
{\bf High twists and the NNLO QCD corrections in DIS}
\vspace{1cm}

{\bf S.~I.~Alekhin}
\vspace{0.1in}

{\baselineskip=14pt Institute for High Energy Physics, 142281 Protvino, Russia}
\end{center}
\begin{abstract}
We discuss interplay between the high-twist (HT) terms in the 
operator-product expansion and the next-to-next-to-leading-order (NNLO)
QCD corrections to the deep-inelastic-scattering structure 
functions in analysis of the high-precision data for charged leptons.
Under account of the NNLO corrections the observed HT terms change within
their experimental errors only and do not vanish in the NNLO.
\end{abstract}

The high-twist (HT) contribution to the deep-inelastic-scattering (DIS)
structure functions arising in the operator-product expansion (OPE)
obey the power-like dependence on the 
momentum transferred $Q$. This $Q$-dependence of the HT terms 
differs from the 
logarithmic-like $Q$-dependence typical for the leading-twist (LT) terms.
Nevertheless the problem of separation of the HT and LT terms
is discussed for many years. This problem was recognized in  
Ref.\cite{Abbott:1980as} and was confirmed with  
the observation that the HT terms decrease under account of the 
NLO QCD corrections to the leading-twist (LT) terms 
in the analysis of DIS data.
Later the decrease of the HT terms under account of the 
NNLO corrections was observed in the 
analysis~\cite{Kataev:1999bp} of $\nu N$ DIS data~\cite{Seligman:mc}.
This interplay between the higher-order (HO) corrections 
and the HT terms is ascribed to both the HO corrections and the HT terms 
are largest at small $Q$ at that the $Q$-dependence of HO
corrections is similar to the $Q$-dependence 
of HT terms ($\alpha_s^n(Q) \sim 1/Q^m$).
Meanwhile interplay of the HO corrections and the HT terms  
is possible only if 
statistical accuracy of the analyzed data is insufficient to 
disentangle the log-type and power-like contributions to the
structure functions (evidently this is the case for 
the data of Ref.~\cite{Seligman:mc}).  In order to study this 
interplay more accurately and clarify the magnitude of HT terms 
we performed the analysis of data on DIS of charged leptons, which 
are by order of magnitude more precise than the neutrino data.

\begin{figure}[t]
\centerline{\epsfig{file=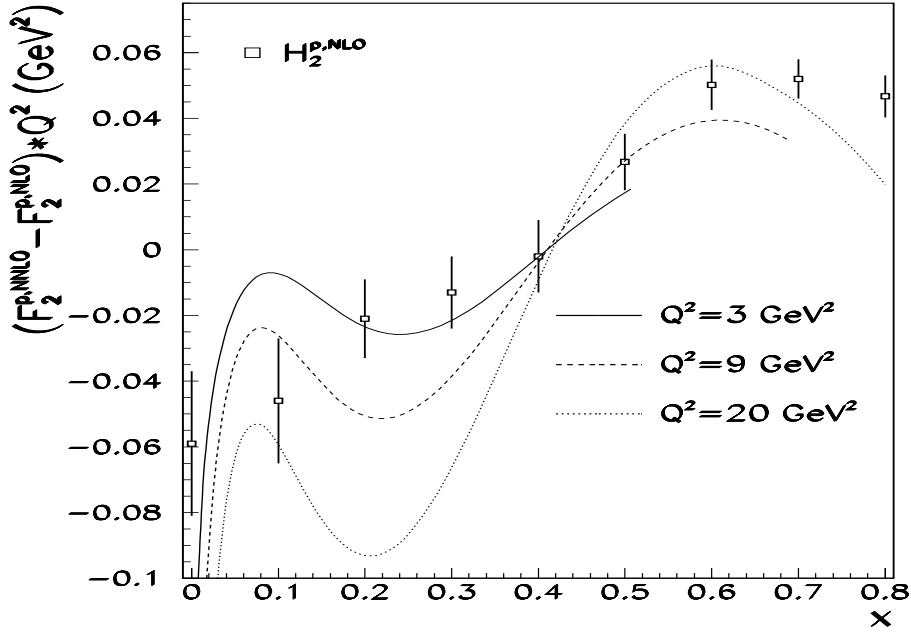,width=14cm,height=9cm}}
\caption{Shown is comparison of the NNLO corrections 
and the HT contribution to $F_2^p$ obtained in the NLO fit.}
\label{fig:dellt}
\end{figure}
\begin{figure}[h]
\centerline{\epsfig{file=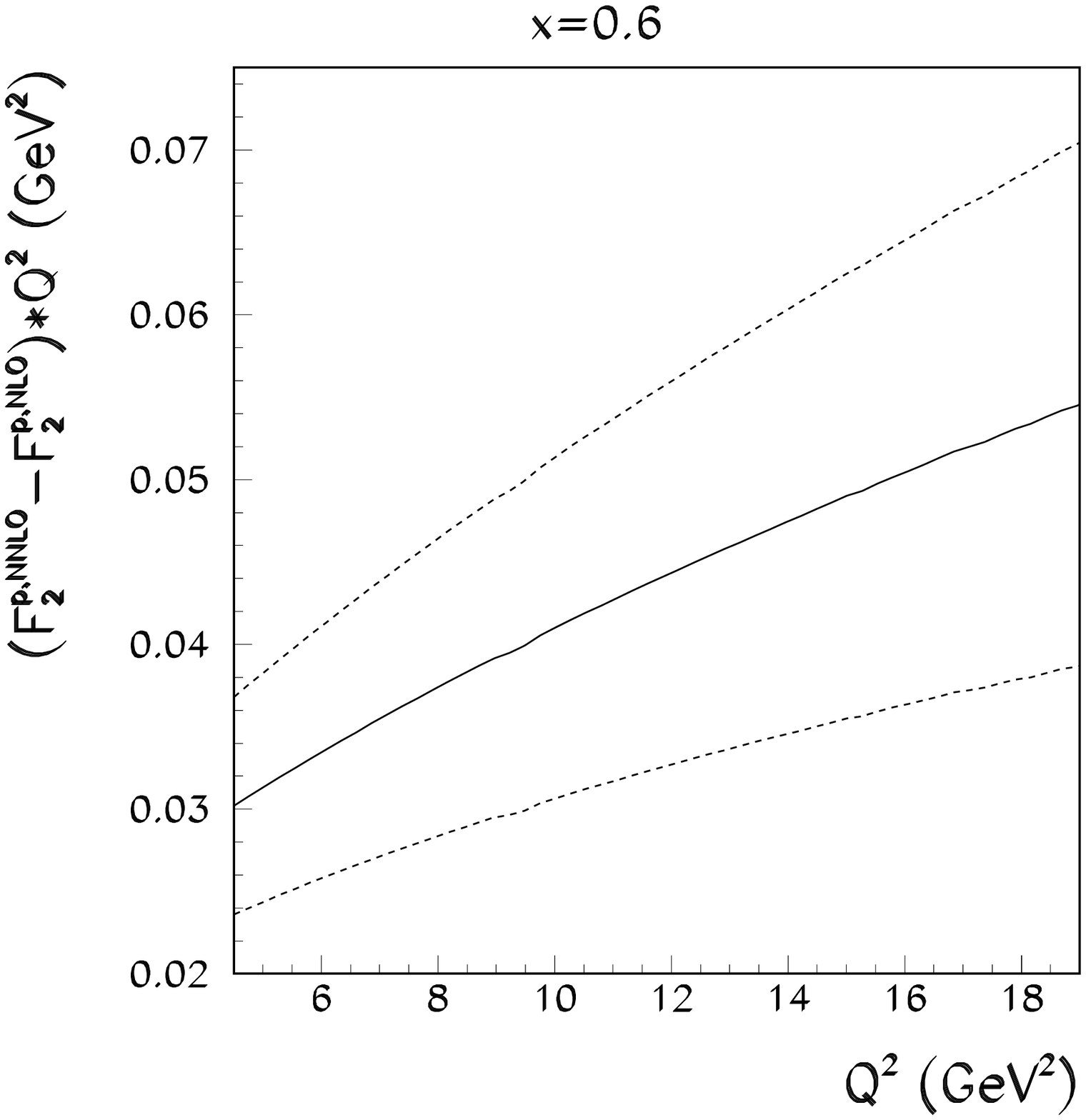,width=14cm,height=10cm}}
\caption{The $Q$-dependence of NNLO correction to $F_2^p$ (full line)
with the experimental errors bands for $F_2^p$ (dashes).}  
\label{fig:delltq}
\end{figure}

We use in the analysis 2274 data points on the cross sections of
DIS scattering of the charged leptons off the proton and deuterium 
targets~\cite{Whitlow:1992uw} 
with $Q^2=2.5 \div 300~$GeV$^2$ and $x=4\cdot 10^{-5}\div 0.75$ in 
order to suppress the theoretical uncertainties which rapidly rise 
at small $Q$ and large $x$. The data were described using  
OPE expansion of the structure functions $F_{2,\rm L}$
$$
F_{2,\rm L}(x,Q)=F_{2,\rm L}^{\rm LT,TMC}(x,Q)+\frac{H_{2,\rm L}(x)}{Q^2}
$$ 
with account of the twist-4 contributions, where the functions 
$H(x)$ are parameterized in the piece-linear form and 
$F_{2,\rm L}^{\rm LT,TMC}$ are the LT terms with account of the
target-mass corrections. To provide self-consistency of the analysis 
the LT terms were calculated using the  
parton distributions fitted simultaneously with the HT terms and 
the value of strong coupling constant $\alpha_{\rm s}$.
The QCD evolution of parton distributions was performed in 
the $\overline{MS}$ scheme with the fixed number of flavors equal 
to 3. The QCD evolution equations with the NNLO 
corrections taken into account using the 
parameterizations of Ref.\cite{vanNeerven:2000wp}
were integrated numerically in the $x$-space.
The code used for the integration was checked against Les Houches 
benchmark~\cite{Giele:2002hx}
and demonstrated precision much better than accuracy of the data used in  
analysis. 

\begin{figure}[t]
\centerline{\epsfig{file=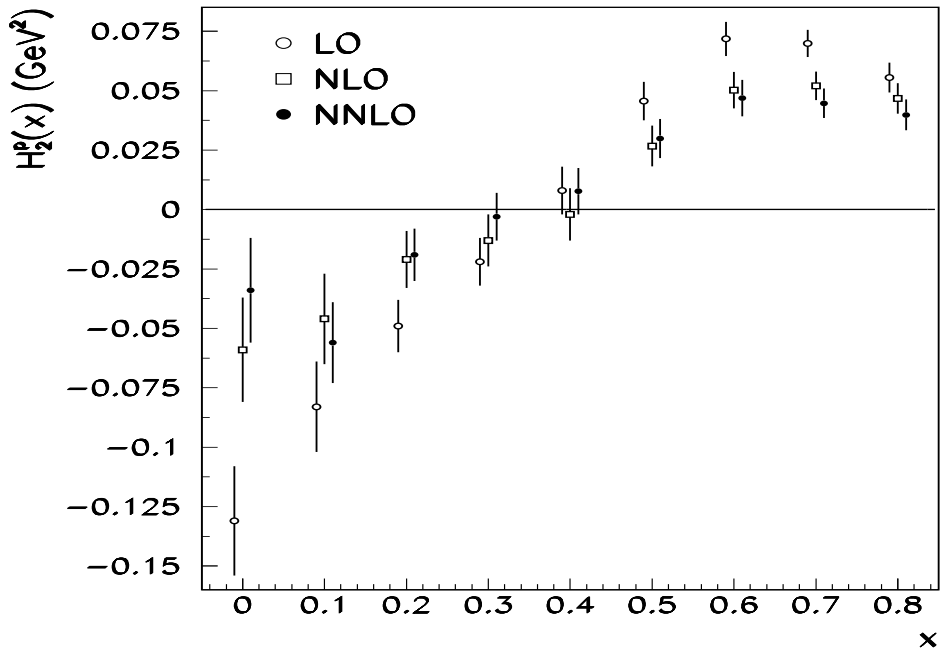,width=14cm,height=10cm}}
\caption{The values of HT terms in $F_2^p$ obtained in 
the LO, NLO, and NNLO fits.}
\label{fig:h2p}
\end{figure}

\begin{figure}[t]
\centerline{\epsfig{file=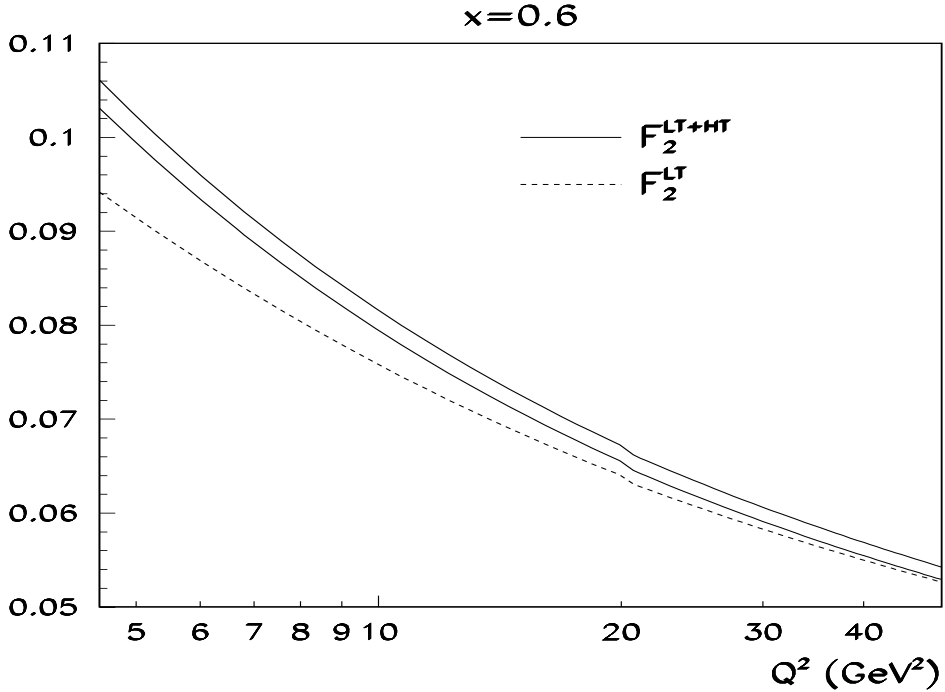,width=14cm,height=10cm}}
\caption{Relative magnitude of the HT and LT terms in $F_2^p$.}
\label{fig:res}
\end{figure}

The HT contribution to the proton structure function $F_2^p$ 
obtained in the NLO approximation is compared to the NNLO correction
to $F_2^p$ in Fig.\ref{fig:dellt}.
The magnitude of $H_2^{p,\rm NLO}$ clearly deviates off 0 within 
the experimental uncertainties, which are combinations of the   
statistical, point-to point correlated systematical, and normalization 
errors in data
accounted in the analysis using the covariance matrix approach.
Value of the NNLO correction is comparable to $H_2^{p,\rm NLO}$,
but its $Q$-dependence does not fit to the 
$1/Q^2$ behavior (see Fig.\ref{fig:delltq}) and
transfer of the NNLO correction to the shift in the
HT term is possible only with changing the quality of fit.
Alternatively, if effect of the NNLO correction is entirely
compensated by the shift in other fitted
parameters (e.g. in value of $\alpha_{\rm s}$), 
quality of the fit would remain the same at that possible  
shift in the HT term is at the level of the fluctuations in data.
The observed shift of $H_2^p$ from the NLO to the NNLO
is of the scale of the experimental errors (see Fig.\ref{fig:h2p}), 
i.e. is not statistically significant, while the value of 
$\chi^2/NDP=1.11$ both for the NLO and the NNLO.
At the same time quality of the fit worsens from the NLO to  
the LO ($\chi^2/NDP=1.20$ in the LO) and this is accompanied by the 
statistically significant shift of $H_2^p$.
Summarizing these observations we conclude that 
poor description of the data in the LO does lead to generation of a fake 
contribution to HT terms, but this contribution essentially decreases
in the NLO (and even more in the NNLO) getting smaller than  
fluctuations in the existing data. 
Remaining deviation of $H_2^p$ off zero can be considered as a  
genuine HT contribution to $F_2^p$.
This contribution is maximal at $x\sim 0.6$. In this region of $x$
the HT term is larger than the experimental error in $F_2^p$
up to $Q^2 \sim 20~{\rm GeV}^2$ (see Fig.\ref{fig:res}) rising 
for small hadronic invariant mass~\cite{Liuti:2001qk}.
This is in disagreement with the results of Ref.\cite{Yang:1999xg},
but the comparison with that analysis is difficult
since it was based on the 
parameterization of HT terms within the infrared renormalon model.

\begin{figure}[t]
\centerline{\epsfig{file=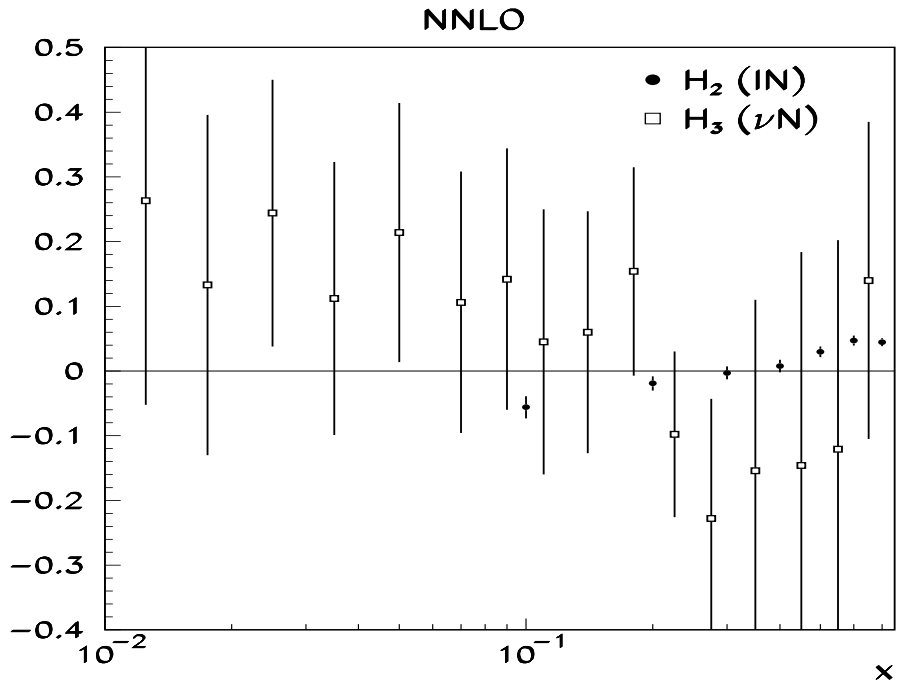,width=14cm,height=10cm}}
\caption{Comparison of the HT terms in the structure functions 
extracted from the $lN$ (our fit) and $\nu N$ DIS data~\cite{Kataev:1999bp}.}
\label{fig:ktf}
\end{figure}

The errors in HT contribution to the structure functions $xF_3^{\nu N}$
obtained in Ref.\cite{Kataev:1999bp}
are much larger than the errors in $H_2^p$ obtained in our analysis
(see Fig.\ref{fig:ktf}). Since the scales of HT terms 
would be similar in both cases, this means
that the precision of existing neutrino data is quite  
insufficient for the quantitative estimation of the 
HT terms. In particular, to clarify the conclusion of Ref.\cite{Kataev:1999bp}
about vanishing the HT terms in the NNLO the 
experiments with luminosity typical for the proposed neutrino factories are 
required~\cite{Mangano:2001mj}.

In conclusion, we observe that in the analysis of $l N$ DIS data  
the higher-twist terms in structure functions 
do not vanish even with account of the NNLO QCD corrections. 
The HT contribution to structure function $F_2$
extracted from the comparison to data
is maximal at $x \sim 0.6$. Its relative error   
is $\sim 20\%$ in this region that allows for conclusive
comparison with different theoretical models.
Account of the HT terms in the analysis of existing DIS data 
is very important because they give non-negligible effect up 
to $Q^2 \sim 20~{\rm GeV}^2$.


\begin{thebibliography}{99}

\bibitem{Abbott:1980as}
L.~F.~Abbott and R.~M.~Barnett,
Annals Phys.\  {\bf 125}, 276 (1980);\\
L.~F.~Abbott, W.~B.~Atwood and R.~M.~Barnett,
Phys.\ Rev.\  {\bf D22}, 582 (1980).

\bibitem{Kataev:1999bp}
A.~L.~Kataev, G.~Parente and A.~V.~Sidorov,
Nucl.\ Phys.\ B {\bf 573} (2000) 405
[arXiv:hep-ph/9905310].

\bibitem{Seligman:mc}
W.~G.~Seligman {\it et al.},
Phys.\ Rev.\ Lett.\  {\bf 79} (1997) 1213.

\bibitem{Whitlow:1992uw}
L.~W.~Whitlow, E.~M.~Riordan, S.~Dasu, S.~Rock and A.~Bodek,
Phys.\ Lett.\  {\bf B282}, 475 (1992);
A.~C.~Benvenuti {\it et al.}  [BCDMS Collaboration],
Phys.\ Lett.\  {\bf B223} (1989) 485;
A.~C.~Benvenuti {\it et al.}  [BCDMS Collaboration],
Phys.\ Lett.\  {\bf B237} (1990) 592;
M.~Arneodo {\it et al.}  [New Muon Collaboration],
Nucl.\ Phys.\  {\bf B483} (1997) 3
[hep-ph/9610231];
C.~Adloff {\it et al.}  [H1 Collaboration],
Eur.\ Phys.\ J.\ C {\bf 21}, 33 (2001)
[arXiv:hep-ex/0012053];
S.~Chekanov {\it et al.}  [ZEUS Collaboration],
Eur.\ Phys.\ J.\ C {\bf 21}, 443 (2001)
[arXiv:hep-ex/0105090].

\bibitem{vanNeerven:2000wp}
W.~L.~van Neerven and A.~Vogt,
Phys.\ Lett.\ B {\bf 490}, 111 (2000)
[arXiv:hep-ph/0007362].

\bibitem{Giele:2002hx}
W.~Giele {\it et al.},
arXiv:hep-ph/0204316.

\bibitem{Liuti:2001qk}
S.~Liuti, R.~Ent, C.~E.~Keppel and I.~Niculescu,
Phys.\ Rev.\ Lett.\  {\bf 89} (2002) 162001
[arXiv:hep-ph/0111063].

\bibitem{Yang:1999xg}
U.~K.~Yang and A.~Bodek,
Eur.\ Phys.\ J.\ C {\bf 13} (2000) 241
[arXiv:hep-ex/9908058].

\bibitem{Mangano:2001mj}
M.~L.~Mangano {\it et al.},
arXiv:hep-ph/0105155.

\end{thebibliography}
\end{document}